\documentclass[aps,prl,twocolumn, superscriptaddress, floatfix]{revtex4}
\usepackage{graphicx,amsfonts,amsmath,color,epsfig,epstopdf, ulem}
\newcommand{\SU}[1]{\ensuremath{\mathrm{SU}( #1 )}}

\newcommand{\SO}[1]{\ensuremath{\mathrm{SO}( #1 )}}

\newcommand{\SpR}[1]{\ensuremath{\mathrm{Sp}( #1,\mathbb{R} )}}
\newcommand{\ket}[1]{\ensuremath{\left| #1 \right\rangle}}

\newcommand{\RedME}[3]{\ensuremath{\langle #1 \| #2 \| #3 \rangle}}

\newcommand{\betb}{\begin{tabular}{p{4.0cm}p{9.0cm}}}
\newcommand{\entb}{\end{tabular}}

\newcommand{\ho}{\ensuremath{\hbar\Omega}}

\newcommand{\Nmax}[2]{\ensuremath{\langle#1\rangle #2}}

\begin{document}

\title{Electron-scattering form factors for $^6$Li in the  {\it ab initio} symmetry-guided framework}
\author{T. Dytrych}
\affiliation{Department of Physics and Astronomy, Louisiana State University, Baton Rouge, LA 70803, USA}
\author{A. C. Hayes}
\affiliation{Theoretical Division, Los Alamos National Laboratory, Los Alamos, New Mexico 87545, USA}
\author{K. D. Launey}
\affiliation{Department of Physics and Astronomy, Louisiana State University, Baton Rouge, LA 70803, USA}
\author{J. P. Draayer}
\affiliation{Department of Physics and Astronomy, Louisiana State University, Baton Rouge, LA 70803, USA}
\author{P. Maris}
\affiliation{Department of Physics and Astronomy, Iowa State University, Ames, IA 50011, USA}
\author{J. P. Vary}
\affiliation{Department of Physics and Astronomy, Iowa State University, Ames, IA 50011, USA}
\author{D. Langr}
\affiliation{Faculty of Information Technology, Czech Technical University, Prague 16000, Czech Republic}
\affiliation{Aerospace Research and Test Establishment, Prague 19905, Czech Republic}
\author{T. Oberhuber}
\affiliation{Faculty of Nuclear Sciences and Physical Engineering, Czech Technical University, Prague 11519, Czech Republic}

\begin{abstract}
We present an \textit{ab initio} symmetry-adapted no-core shell-model description for $^6$Li. We study the structure of the  ground state of $^6$Li and the impact of the symmetry-guided space selection on  the charge density components for this state in momentum space, including the effect of  higher shells. We accomplish this by  investigating the electron scattering charge form factor for momentum transfers up to $q \sim 4$ fm$^{-1}$.
We demonstrate that  this symmetry-adapted framework can achieve significantly reduced dimensions for equivalent large shell-model spaces while retaining the accuracy of the form factor for   any momentum transfer.
These new results confirm the previous outcomes  for selected spectroscopy observables in light nuclei, such as binding energies, excitation energies, electromagnetic moments, $E2$ and $M1$ reduced transition probabilities, as well as point-nucleon matter  rms radii. 
\end{abstract}


\maketitle
\section{ Introduction}
The symmetry-adapted no-core shell model  (SA-NCSM) \cite{DytrychLMCDVL_PRL12} has been recently developed and designed to provide nuclear structure descriptions by using a new, symmetry-adapted and physically relevant many-particle basis. The model has been employed to unveil the emergence of a simple orderly pattern in  nuclear dynamics, for the first time,  in an {\it ab initio} framework (that is, from first principles), without {\it a priori} symmetry constraints. This highly structured formation is associated with an  approximate symmetry in low-lying nuclear states  that has been earlier suggested and linked to the symplectic \SpR{3} group and its embedded \SU{3} group  \cite{BohrMottelson69,Elliott,Hecht71,Sp3R1,RoweRPP,DraayerWR84,BahriR00,Vargas01,DytrychSBDV07}. The pattern favors  low intrinsic spin together with large deformation and symplectic excitations thereof. This provides a strategy for determining the nature of bound states of nuclei in terms of a relatively small fraction of the possible configurations. Consequently, we may 
extend the reach of \textit{ab initio} approaches 
\cite{NCSM,BarrettNV13,MarisVN13,GFMC,CCM,
NCSMstudies2,NCSMreactions2,BognerFMPSV08,RothLCBN11,EpelbaumKLM11}  to explore ultra-large model spaces for a description of heavier nuclei and highly deformed structures together with the associated rotations. We have demonstrated that the SA-NCSM reduces the  model space through a very structured selection, based on symmetry considerations, to physically relevant subspaces without compromising the accuracy of the {\it ab initio} NCSM approach \cite{DytrychLMCDVL_PRL12}. 

In this paper, we focus on elastic $(e,\, e^\prime)$  scattering charge form factors for the ground state of $^6$Li  and show that the SA-NCSM model with a symmetry-guided space selection provides a description of 
the form factors equivalent to the ones obtained in the corresponding complete space. This holds for any momentum transfer, from low  $ q \lesssim 1$ fm$^{-1}$ through intermediate (up to $3$ fm$^{-1}$), and above (shown here up to $q \sim 4$ fm$^{-1}$).
While results show that theoretical form factors are reasonably trending towards experiment, the $^6$Li  charge radius is not completely converged, so high-precision comparisons with experiment remain for future work. Nevertheless, the results presented here show, 
for the first time, that the calculated ground-state ({\it gs}) one-body charge density components in momentum space, including the contribution from excitations to higher harmonic oscillator (HO) shells, 
is properly taken into account in  selected spaces guided by \SpR{3} and  \SU{3} symmetry considerations  (similarly, for low-lying eigenstates of the {\it gs} rotational band). This, together with earlier SA-NCSM findings for observables such as binding energies, excitation energies, electromagnetic moments, $E2$ and $M1$ reduced transition probabilities, as well as point-nucleon matter  rms radii for selected states \cite{DytrychLMCDVL_PRL12}, confirms the validity of the SA-NCSM concept.

The significance of electron scattering form factors studies  stems from their ability to provide a 
probe of the  structure of  the wavefunctions. For example, Ref. \cite{HayesNV03} studied inelastic scattering form factors and cross sections to discern important spin flip components in $^{12}$C wavefunctions that were sensitive to three-nucleon interactions. In this paper,
we examine the longitudinal form factor (C0) for scattering off the ground state of $^6$Li that is a Fourier transform of the ground-state charge density. 
The C0 form factors provide an indication on how well  nuclear structure calculations reproduce the different lower- and higher-momentum transfer components of the nuclear charge density.  This, in turn, can reveal important underlying physics responsible for achieving convergence of the moments of the charge density starting with the rms radius.
\begin{figure*}[th]
\includegraphics[width=1\textwidth]{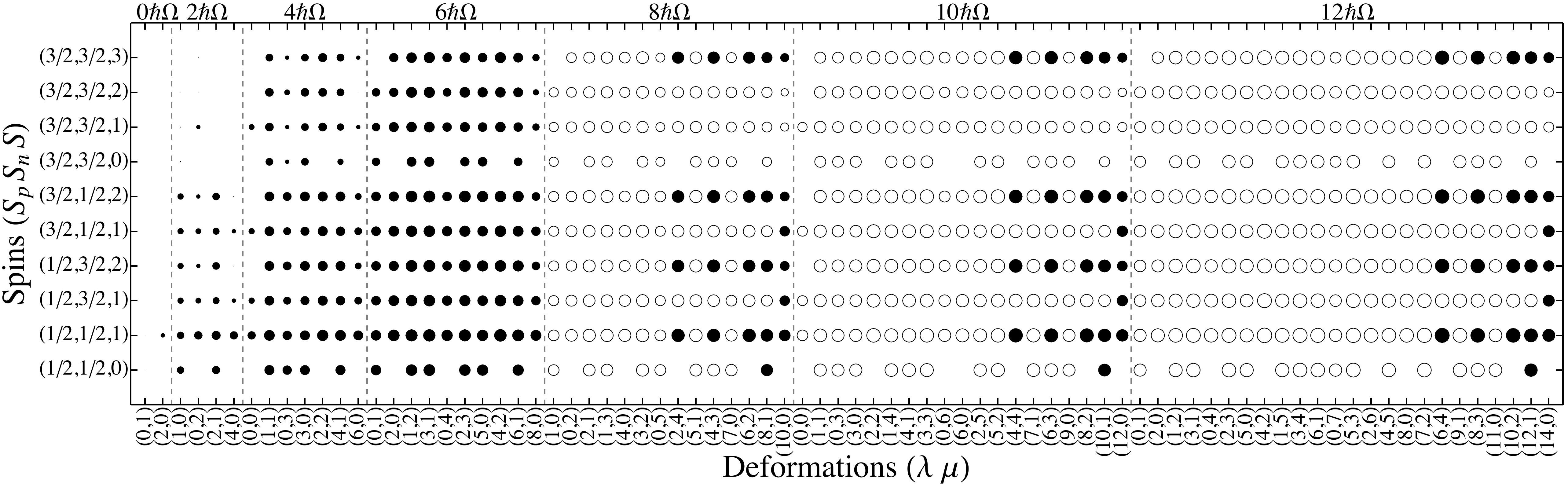}
\caption
{
$N_{\max}=12$ model space (all circles, filled and unfilled) specified by the proton, neutron, and total intrinsic
	($S_{p},\, S_{n},\, S$) spin  values (vertical axis) across the Pauli-allowed deformation-related $(\lambda\,\mu)$
	values (horizontal axis) for the $1^{+}$ ground  state of $^{6}$Li. The selected \Nmax{6}{12} SA-NCSM model space is shown by filled circles  and it includes the complete space up through the 6\ho~subspace, while the 8\ho, 10\ho, and 12\ho~subspaces are selected based on symmetry considerations favoring large deformation. The SA-NCSM results in a model space, for which all circles are filled up through the 12\ho~subspace, coincide with the $N_{\max}=12$ NCSM results.
}
\label{modelSpace_Li6}
\end{figure*}

The charge form factors are calculated in the first-order plane-wave Born approximation. In all $(e,\, e^\prime)$ calculations presented here we use bare interactions, namely, the realistic nucleon-nucleon ($NN$) NNLO$_{\rm opt}$ \cite{Ekstrom13} and JISP16 \cite{ShirokovMZVW04} (with similar results obtained for N$^3$LO \cite{EntemM03}). The use of bare interactions, and not effective interactions in smaller model spaces, implies that operators used to calculate form factors does not have to be renormalized.
In addition, charge form factors are calculated using the one-body charge density multipole operator, while contributions from two-body
charge operators and/or relativistic corrections are not considered, as they are known to be negligible  for charge form factors  for momenta up to about $q \approx 2$ fm$^{-1}$ \cite{HayesK13}. Our calculated form factors  have no center-of-mass (CM) contribution and are further adjusted to account for the finite proton size. 

\section{Symmetry-guided framework and electron scattering form factors}
A detailed description of the {\it ab initio} symmetry-adapted no-core shell model  (SA-NCSM) has been presented, e.g., in Refs. \cite{DraayerDLL12,Dytrych13_C12}. 
The SA-NCSM adopts the first-principle concept and is a no-core shell model (NCSM) carried forward in an \SU{3}-coupled scheme \cite{Elliott}. 
The conventional NCSM \cite{NCSM} calculations are carried out in many-particle
basis of Slater determinants (SD)
built on HO single-particle states  characterized by the $\hbar\Omega$ oscillator frequency (or equivalently, the oscillator length $b=\sqrt{\hbar/m\Omega}$). The model space is spanned by nuclear configurations of fixed parity,
consistent with the Pauli principle, and truncated by a cutoff
$N_{\max}$. The $N_{\max}$ cutoff is defined as the maximum number of HO quanta
allowed in a many-particle state above the minimum for a given nucleus.
\begin{figure*}[th]
\includegraphics[width=0.49\textwidth]{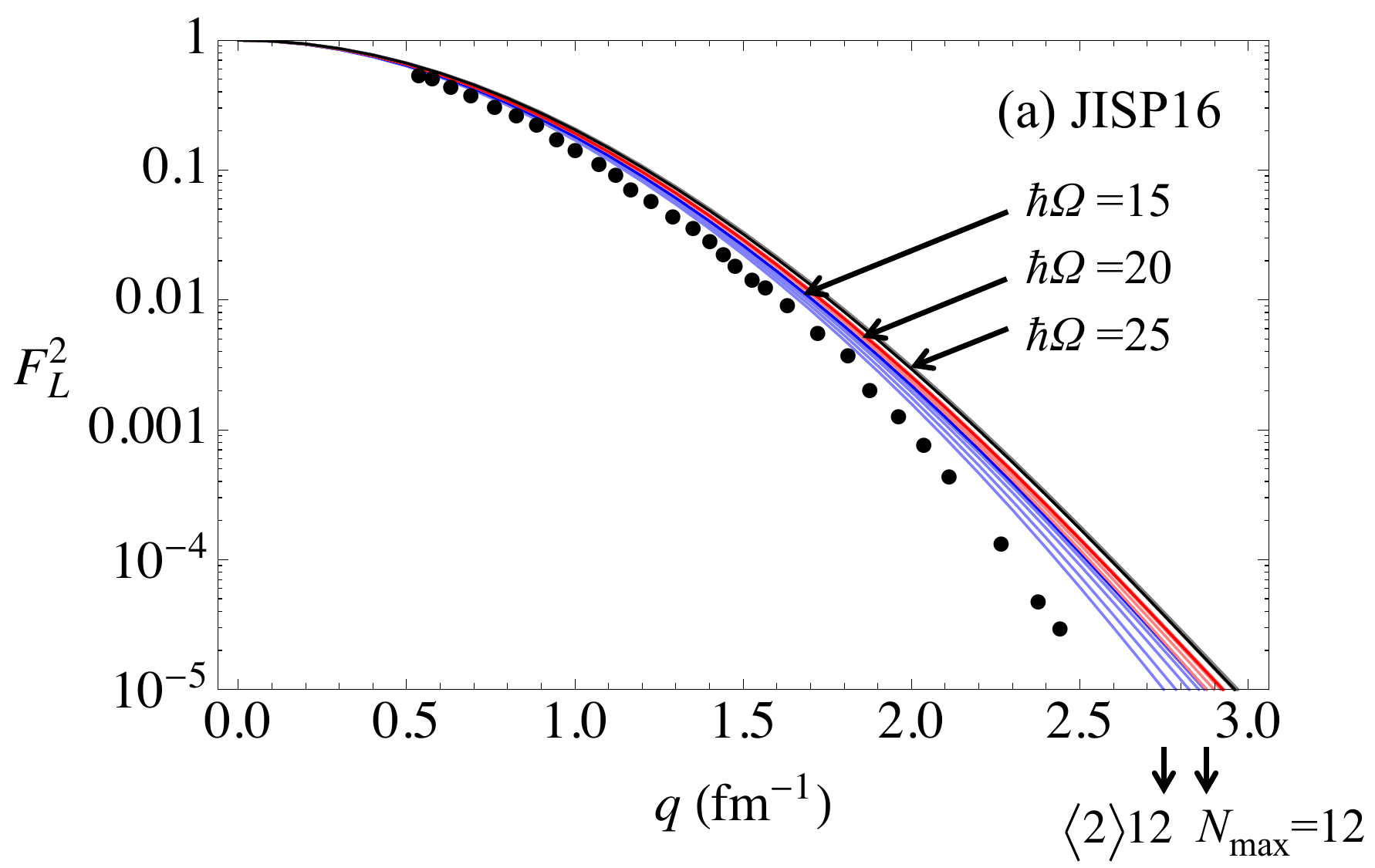}
\includegraphics[width=0.49\textwidth]{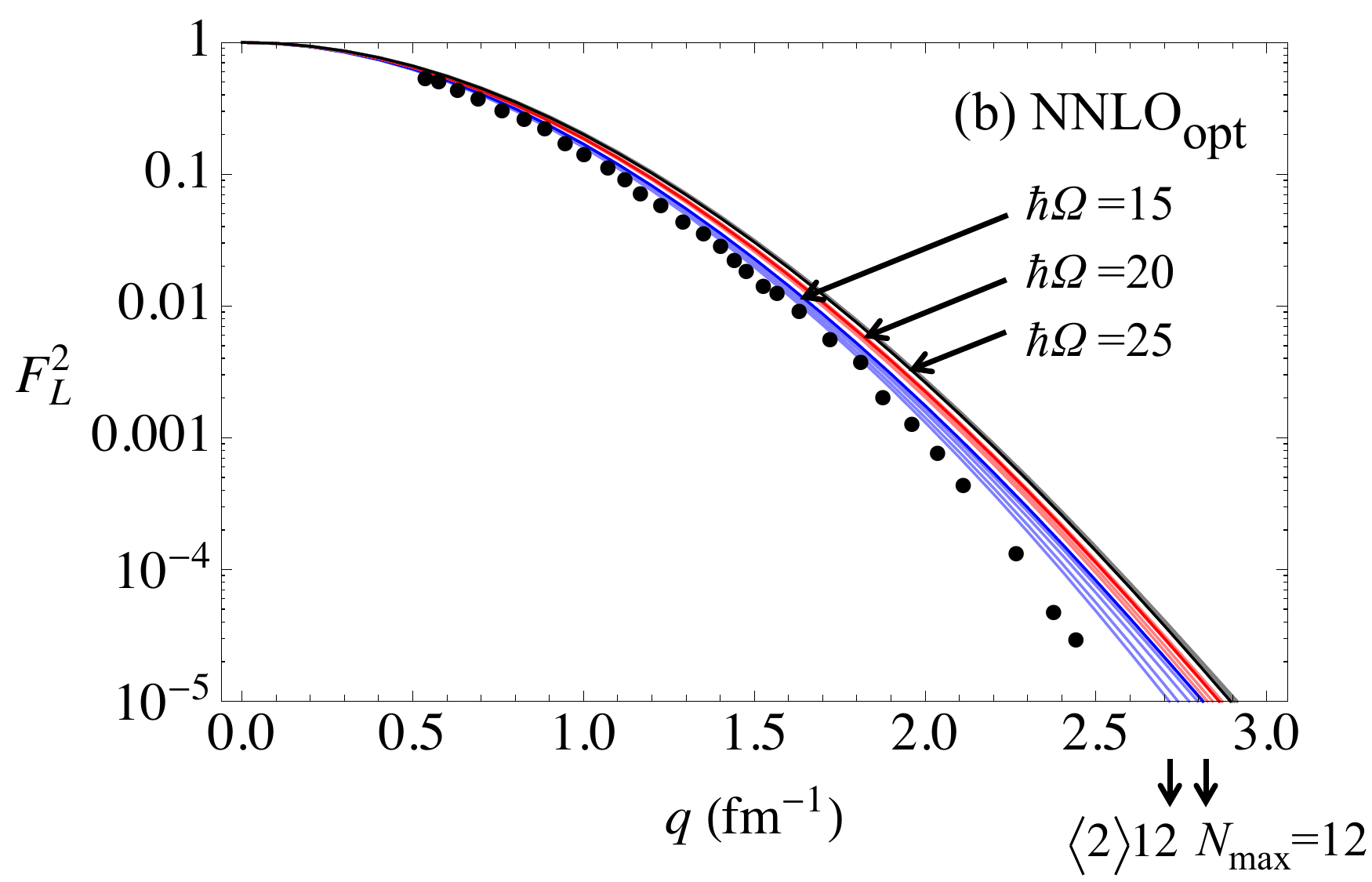} 
\caption
{
(Color online) Longitudinal $C0$ electron scattering form factors $F_L^2$ (the $L_0 = 0$ term in Eq. (\ref{F2L}) with a procedure discussed in the text to produce a translationally invariant form factor)  for the SA-NCSM $1^{+}$ ground  state of $^{6}$Li calculated in the complete $N_{\max}=12$ space (darker colors) and the \SU{3}-selected spaces, \Nmax{2}{12}, \Nmax{4}{12}, \Nmax{6}{12}, \Nmax{8}{12}, and \Nmax{10}{12} (lighter colors), for $\hbar\Omega=15$ MeV or $b=1.66$ fm (blue), $\hbar\Omega=20$ MeV or $b=1.44$ fm (red), and $\hbar\Omega=25$ MeV or $b=1.29$ fm (black) for (a) the bare JISP16 interaction, as well as for (b) the bare  NNLO$_{\rm opt}$ interaction. Experimental data are taken from Ref. \cite{LiSWY71}.
}
\label{FLC0_Li6}
\end{figure*}

The many-particle basis states of the SA-NCSM for a given $N_{\max} $ are constructed in the proton-neutron
formalism and are  labeled by the
quantum numbers $(\lambda\,\mu)\kappa L$ of the \SU{3}$_{(\lambda\,\mu)}\underset{\kappa}{\supset}$\SO{3}$_L$ group chain, together with proton, neutron, and total
intrinsic spins $S_{p}$, $S_{n}$, and $S$ of the complementary \SU{2} spin group. The label $\kappa$ distinguishes multiple occurrences 
of the same $L$ value in the parent irrep $(\lambda\,\mu)$. The orbital angular momentum $L$
is coupled with $S$ to the total angular momentum $J$ with a projection
$M_{J}$. Each basis state in this scheme is labeled schematically as
$ |\vec{\gamma}\, N(\lambda\,\mu)\kappa L;(S_{p}S_{n})S;J M_{J}\rangle$, where $N$ is the total number of HO excitation quanta and 
 $\vec{\gamma}$ denotes additional quantum numbers needed to
distinguish among configurations carrying the same $N(\lambda\,\mu)$ and
($S_{p}S_{n})S$ labels. The organization of the model space allows the full space to be down-selected to the physically relevant subspace. 

The significance of  the \SU{3} group for a microscopic description of the nuclear dynamics can be seen from the fact that it is the symmetry group of the established Elliott model~\cite{Elliott}, and a subgroup of the \SpR{3}, the underpinning symmetry of the successful microscopic symplectic model~\cite{Sp3R1}. 

The charge form factors are calculated in the first-order plane-wave Born approximation. They are derived using the formalism and an extension of the computer code developed by Lee \cite{Lee}, described in detail in Ref. \cite{HayesK13}, as well as using an \SU{3}-based apparatus \cite{RochfordD92,EscherD99} for calculating charge and current
density distributions in terms of the shell-model one-body
density matrix elements (OBDMEs) and the single-particle matrix elements of
the associated electromagnetic operators. We calculate the OBDMEs using wavefunctions obtained in the {\it ab initio} SA-NCSM in complete $N_{\rm max}$ spaces  or selected $\langle N^{\bot}_{\max} \rangle N_{\max}$ spaces.  An $\langle N^{\bot}_{\max} \rangle N_{\max}$ model space includes the complete basis up through  $N^{\bot}_{\max}$ along with selected $(\lambda\,\mu)$ and $(S_p S_n S)$ configurations
beyond $N^{\bot}_{\max}$ up through $N_{\max}$ (see Fig. \ref{modelSpace_Li6} for a \Nmax{6}{12} model space).

In the present analysis, we use the  SA-NCSM with  two realistic $NN$ interactions, the
bare JISP16~\cite{ShirokovMZVW04} and NNLO$_{\rm opt}$~\cite{Ekstrom13} potentials.
 The Coulomb interaction is added along with the  $NN$ interaction, together with a Lawson term for elimination of spurious
center-of-mass excitations.
We present results for $N_{\max}=12$, as this model space is found sufficient to  achieve convergence of the $^{6}$Li  {\it gs} energy --  e.g., for $\ho=20$ MeV, it is within $0.54$ MeV of the extrapolated result of $-31.49(6)$  MeV  \cite{CockrellVM12, MarisV13, ShirokovKMV14}.  
Electron-scattering calculations are performed for a range of  $\hbar\Omega=15, 20,$ and $25$ MeV and for several \SU{3}-selected spaces, \Nmax{2}{12}, \Nmax{4}{12}, \Nmax{6}{12}, \Nmax{8}{12}, \Nmax{10}{12}, together with  the complete $N_{\max}=12$ space. The resulting wavefunctions, $\ket{\alpha JM_J}$ (where $\alpha $ distinguishes different eigenstates of given angular momentum $J$), are used to calculate lab-frame (or SD) OBDMEs, 
\begin{equation}
\RedME{\alpha_f J_f}{\{ a^\dagger_{n_1l_1j_1;t_z} \times \tilde a_{n_2l_2j_2; t_z}\}^{J_0}}{\alpha_i J_i},
\label{obdme}
\end{equation} 
where $nlj$ label single-particle HO basis states and $t_z$ is either proton or neutron [$ \tilde a_{nlj; t_z}$ is the annihilation \SU{2} tensor operator, which destroys a proton or neutron in an $nlj$ state]. These matrix elements are utilized to calculate longitudinal form factors for  scattering from an arbitrary initial (``$i$") eigenstate  to an arbitrary final (``$f$") eigenstate as a function of the three-momentum transfer $q=|{\mathbf q}|$:
\begin{equation}
F^2_{L}(q)=\frac{4\pi }{Z^2(2J_i+1)}\sum_{L_0} | \RedME{\alpha_f J_f}{M_{L_0}(q)}{\alpha_i J_i} |^2,
\label{F2L}
\end{equation}
where the sum is restricted to $|J_i-J_f| \le L_0 \le J_i+J_f$ and  $M_{L_0}(q)=\int j_{L_0}(qr)Y^{L_0}_M(\hat r) \rho(\mathbf r)d^3\mathbf r = \sum_{i =1{\rm (protons)}}^A j_{L_0}(qr_i)Y^{L_0}_M(\hat r_i)$ is the charge density multipole operator  given in position operators relative to the CM position operator. As the  $M_{L_0}(q)$ is  a one-body operator, its reduced matrix elements that enter in Eq. (\ref{F2L}) can be expressed in terms of the OBDMEs of Eq. (\ref{obdme}), provided the contribution of the CM component of the wavefunctions is properly removed. That is, the OBDMEs are calculated for shell-model wavefunctions with lab coordinates and a CM  component in the lowest $0s$ state,  and hence, the CM-free $F^2_{L}(q)$ is calculated using the lab-frame OBDMEs of Eq. (\ref{obdme}) multiplied by an overall factor of 
$e^{2\frac{b^2q^2}{4A} }$, which removes the contribution of the CM component to the  $F^2_{L}(q)$ \cite{TassieB58} (see also \cite{RochfordD92}). In addition, the calculated $F^2_{L}(q)$ form factors are adjusted to account for the finite proton size \cite{SimonSBW80}.
For the elastic electron scattering off the $^{6}$Li ground state, the $C0$ form factor, given by $F^2_{L}$ of Eq. (\ref{F2L}) with  $L_0=0$, is calculated for $J_i=1$, $J_f=1$, and $J_0=0$. 
\begin{figure}[th]
\includegraphics[width=0.47\textwidth]{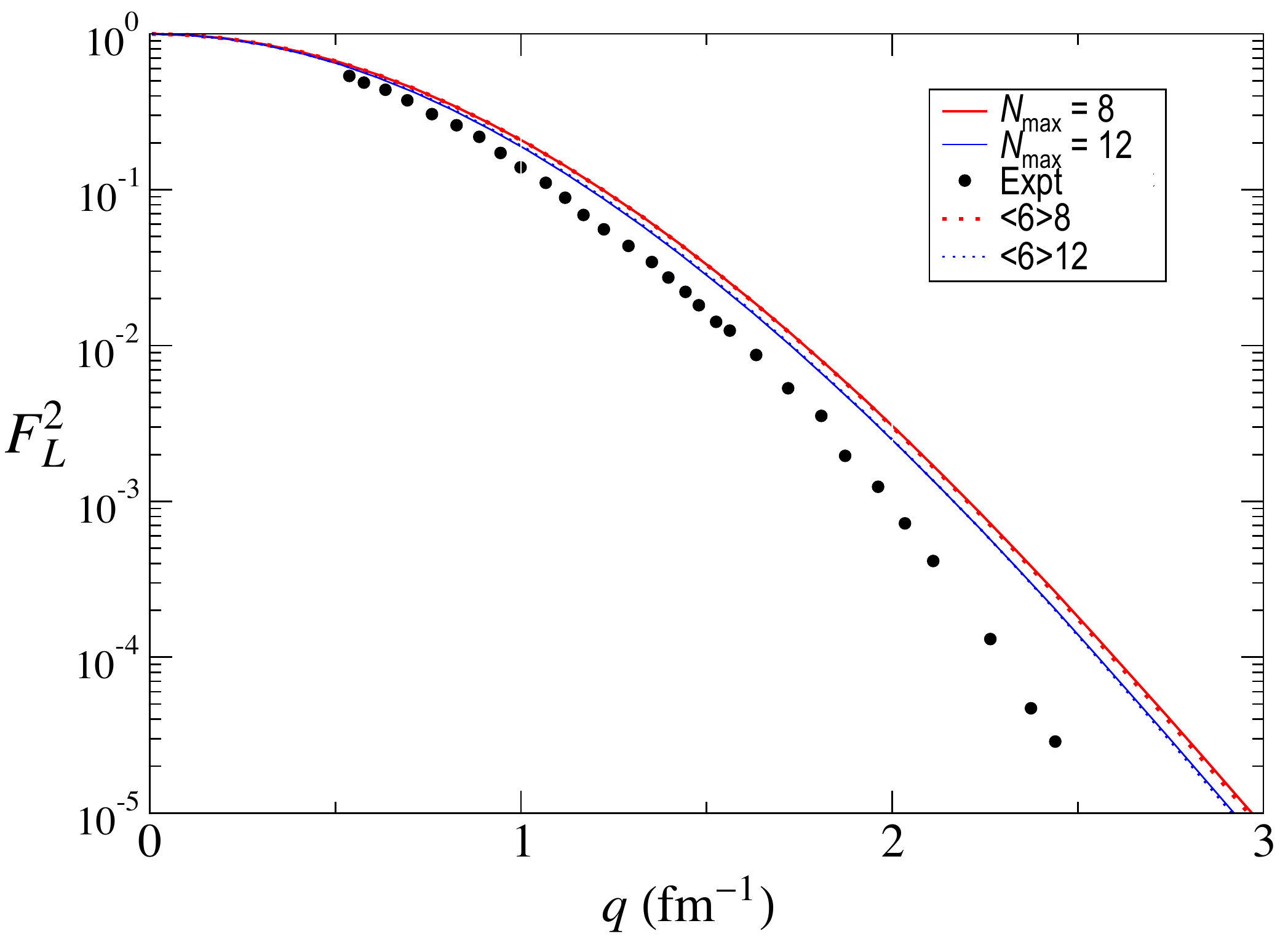}
\caption
{
(Color online) Longitudinal $C0$ electron scattering form factors $F_L^2$ (with a procedure discussed in the text to produce a translationally invariant form factor) for the SA-NCSM $1^{+}$ ground  state of $^{6}$Li calculated  for $\hbar\Omega=20$ MeV or $b= 1.44 $ fm  and with the bare JISP16 interaction. The outcome for the \SU{3}-selected spaces,  \Nmax{6}{8} (red dots) and \Nmax{6}{12} (blue dots), accurately reproduces the corresponding results for the complete $N_{\max}=8$ space (solid, red) and $N_{\max}=12$ space  (solid, blue), with larger-space $N_{\max}=12$ results lying slightly closer to experiment
 \cite{LiSWY71}. 
}
\label{FLC0_Li6vsNmax}
\end{figure}

\section{Results and discussions}
\subsection{Symmetry-guided form factors  in the low- and intermediate-momentum transfer regime}
Longitudinal electron scattering form factors for the ground  state of $^{6}$Li are studied for the bare JISP16 and NNLO$_{\rm opt}$  $NN$ interactions up to $N_{\max}=12$ spaces. An important result is that in all cases, $\Nmax{6}{12}$ selected-space results are found to be almost identical to the $N_{\rm max}=12$ complete-space counterparts in low- and  intermediate- momentum transfer regions (Fig. \ref{FLC0_Li6}), and even above $3$ fm$^{-1}$ (not shown in the figure). This remains valid for various \ho~ values, as well as when different interactions are employed (Figs. \ref{FLC0_Li6}a and \ref{FLC0_Li6}b). It also applies to every state of the $^{6}$Li  $gs$ rotational band, as these states share the same \SU{3} structure \cite{DytrychLMCDVL_PRL12}, but different total orbital momenta, and have very similar OBDMEs. This further confirms the validity of the symmetry-guided concept in the SA-NCSM. Indeed, while we have shown in Ref. \cite{DytrychLMCDVL_PRL12} that  the $N_{\max}=12$ complete-space binding energies, excitation energies, electromagnetic moments, $E2$ and $M1$ reduced transition probabilities, as well as point-nucleon matter  rms radii are accurately reproduced in small selected spaces, the present results indicate that using these selected spaces, that constitute only a fraction of the complete $N_{\rm max}$ model space (about $1$\% for $\Nmax{6}{12}$), 
reproduces, in addition, the $N_{\rm max}=12$ complete-space form factor momentum dependence. In short, model-space selection,  
which is based on a straightforward prescription dictated by the \SpR{3} and  \SU{3} symmetries, 
eliminates many-body basis states that are shown in this study to be also irrelevant for
describing the charge distribution for the $^{6}$Li {\it gs} as revealed by the $C0$ form factor at low/intermediate momentum transfers and above.

Deviations in the form factor as a result of the \SU{3}-based selection of model spaces are found to decrease  for higher \ho~ values  (see Fig. \ref{FLC0_Li6}: the higher the \ho~ value, the narrower the curve). This effect is more prominent for momenta $q > 2$ fm$^{-1}$. The outcome suggests that for high enough \ho~ values, results are almost independent from the model-space selection and, for $\ho=25$ MeV,  the $\Nmax{2}{12}$ form factor already reproduces the $N_{\rm max}=12$ complete-space result. For low \ho~ values, larger $N^{\bot}_{\max}$ spaces ($\Nmax{4}{12}$ or $\Nmax{6}{12}$) appear necessary pointing to a mixing of more deformation/spin configurations within these low-\ho~ spaces. However, while low values, $\ho \lesssim 15$ MeV, are known to require larger model spaces to obtain convergence of the {\it gs} energy, such a mixing at the 4\ho~ and 6\ho~ subspaces is expected to decrease for $N_{\max}>12$. In short, the \SU{3}-based selection of the model space yields  reasonably small deviations in the form factor, especially for $q<2$ fm$^{-1}$ and for $\ho>15$ MeV.
\begin{figure*}[th]
\includegraphics[width=1\textwidth]{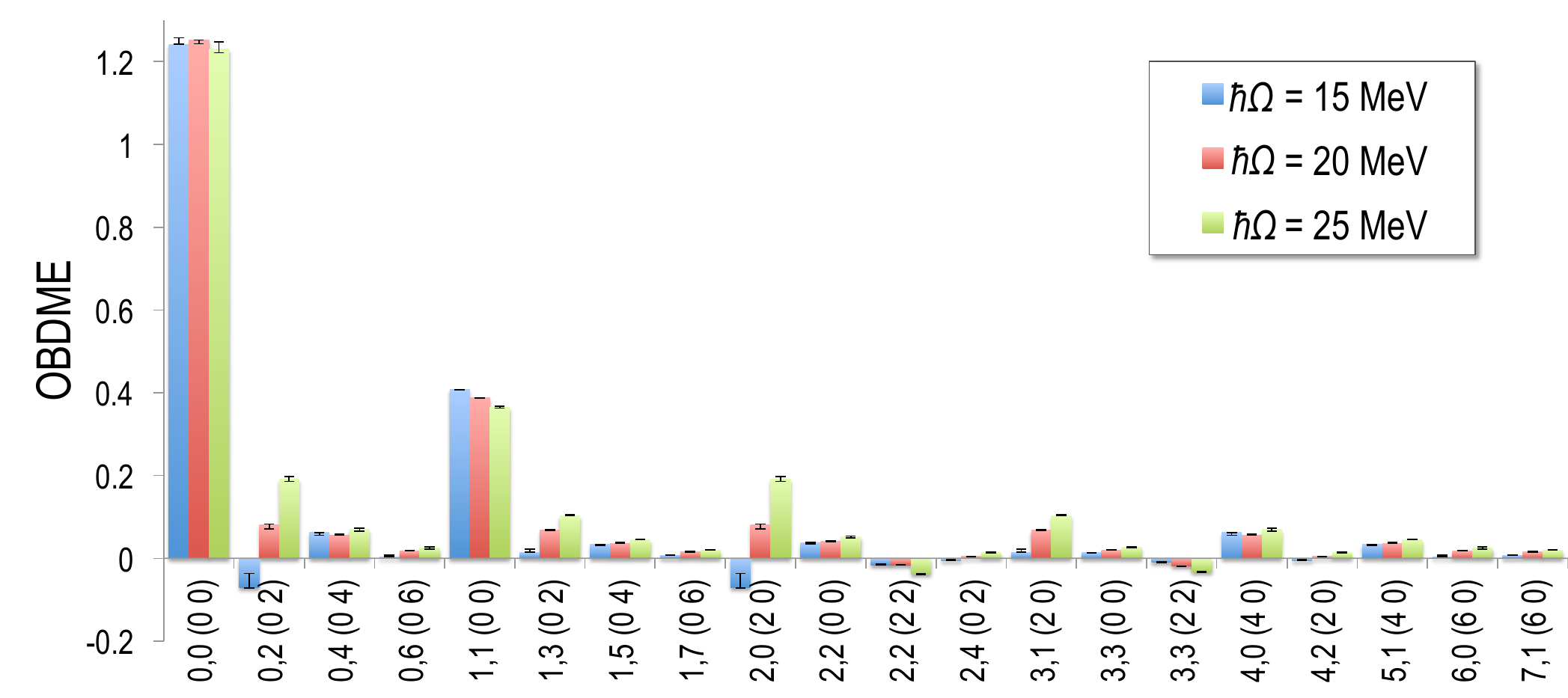}
\caption
{
(Color online) Most dominant OBDMEs (with an absolute value $>0.02$) labeled by $n_1, n_2(\lambda\,\mu)$ for a particle-hole $(n_1)^1n_2^{-1}$ configuration, for the SA-NCSM $1^{+}$ ground  state of
	$^{6}$Li calculated in lab coordinates (with the Lawson term employed in the SA-NCSM calculations ensuring  a $0s$ CM wavefunction component, {\it i.e.},  spurious CM excitations eliminated) and in the $N_{\max}=12$  complete space with the
	JISP16 bare interaction for $\hbar\Omega=15$ MeV (blue, left bars),  20  MeV  (red, middle bars), and 25  MeV  (green, right bars). Error bars are defined by the range from the lowest value to the largest value of each OBDME over the set of \SU{3}-selected spaces.
}
\label{fig:obdmes}
\end{figure*}
\begin{table}[th]
\caption{
Binding energy (BE), excitation energies ($E$), electric quadrupole ($Q$) and
magnetic dipole ($\mu$) moments, as well as point-nucleon proton ($r_p$) and
matter  ($r_m$) rms radii for the three lowest-lying $T=0$ states in $^{6}$Li,
as calculated in the  \Nmax{6}{12} SA-NCSM with the JISP16 $NN$ interaction and
for \ho=20 MeV (taken from Ref.  \cite{DytrychLMCDVL_PRL12}) and compared to
other {\it ab initio} approaches:  the complete $N_{\rm max}=12$ model
space  \cite{DytrychLMCDVL_PRL12} (or NCSM for JISP16 and \ho=20 MeV), as
well as Variational Monte Carlo (VMC) and Green's function Monte Carlo
(GFMC) using the AV18 two-nucleon and Urbana IX three-nucleon interactions
(energies taken from Ref. \cite{WiringaS98}; radii and electromagnetic
moments  taken from Ref.  \cite{PudlinerPCPW97}, without contributions from
two-body currents). Experimental results (Expt.) taken from Ref.
\cite{Tilley02} unless otherwise specified.
}
\begin{center}
\begin{tabular}{l|lllll}
\hline\hline
		&	SA-NCSM 	&	NCSM	 & VMC	&	GFMC	&	Expt.	\\
	\hline
		&		&		&	$1^+_{gs}$	&		&		\\
	BE [MeV]	&	30.445	&	30.951	&	27.0(1)	&	31.2(1)	&	31.99 	\\
	rms $r_p$ [fm]	&	2.112	&	2.125	&	2.46(2) 	&		&	2.43\tablenote{Deduced from the $^{6}$Li charge radius of 2.56(5) fm \cite{LiSWY71}} 	\\
	rms $r_m$ [fm]	&	2.106	&	2.119	&		&		&	2.35(3)\tablenote{From Ref. \cite{Tanihata88}} 	\\
	$Q$ [$e$ fm$^2$]	&	-0.08	&	-0.064	&	-0.33(18)	&		&	-0.0818(17)	\\
	$\mu$ [$\mu_N$]	&	0.839	&	0.838	&	0.828(1)	&		&	0.822 	\\
		&		&		&	$3^+$	&		&		\\
	$E$ [MeV]	&	2.515	&	2.526	&	3.0(1)	&	2.7(3)	&	2.186	\\
	rms $r_m$ [fm]	&	2.044	&	2.063	&		&		&		\\
	$Q$ [$e$ fm$^2$]	&	-3.88	&	-3.965	&		&		&		\\
	$\mu$ [$\mu_N$]	&	1.866	&	1.866	&		&		&		\\
		&		&		&	$2^+$	&		&		\\
	$E$ [MeV]	&	5.303	&	5.066	&	4.4(1)	&	4.4(4)	&	4.312	\\
	rms $r_m$ [fm]	&	2.18	&	2.204	&		&		&		\\
	$Q$ [$e$ fm$^2$]	&	-2.279	&	-2.318	&		&		&		\\	
	$\mu$ [$\mu_N$]	&	1.014	&	0.97	&		& & \\
	\hline\hline
\end{tabular}
\end{center}
\label{tab:cmp_abinitioModels}
\end{table}%

While results using NNLO$_{\rm opt}$ lie slightly closer to experiment, both
interactions show similar patterns with a small dependence on \ho~(Fig.
\ref{FLC0_Li6}).  Furthermore, as one increases $N_{\max}$ (e.g., from
$N_{\max}=8$ to $N_{\max}=12$), SA-NCSM predictions are  reasonably trending
towards experiment, as  illustrated for a  $\Nmax{6}{N_{\max}}$ selected space
and for the reasonable \ho=20 MeV in Fig. \ref{FLC0_Li6vsNmax}.  We note that
the  $N_{\max}=12$ results continue to deviate from the experimental data for
intermediate momenta, especially for $q \gtrsim 2$ fm$^{-1}$. Agreement with
experiment may also depend on including contributions of three-body
interactions in the SA-NCSM calculations and two-body operators in the $F^2_L$. 
The significance of these contributions has been shown in the framework of the
Variational Monte Carlo (VMC) with the AV18 \cite{WiringaSS95} two-nucleon and
Urbana IX \cite{PudlinerPCW95} three-nucleon  interactions \cite{WiringaS98}.
The  low-\ho~ SA-NCSM $F_L^2$ calculations using NNLO$_{\rm opt}$  agree with
the ones of the VMC using AV18/UIX  (without contributions from two-body
currents) for  $q \lesssim 2$ fm$^{-1}$.  The agreement might be a consequence
of the fact that the NNLO$_{\rm opt}$  is designed to minimize the contribution
due to three-nucleon interactions (similarly, for JISP16).  In order to gain
additional insight into the similarities and differences among the {\it ab
initio} results for $^6$Li, we present in Table \ref{tab:cmp_abinitioModels}
the energies, electromagnetic moments, and point-nucleon  rms radii for
selected states in $^{6}$Li, as calculated in the present SA-NCSM approach with
the JISP16 and NNLO$_{\rm opt}$, and in other {\it ab initio} models, such as
the VMC with AV18/UIX and the Green's function Monte Carlo (GFMC) with
AV18/UIX.  The results presented in Table \ref{tab:cmp_abinitioModels} show
good correlations among the different models with, perhaps, the exception of
the smaller rms radii obtained with JISP16 and the larger magnitude of the
electric quadrupole moment obtained with the VMC.  We note that the VMC with
AV18/UIX has shown that two-body currents become significant for $C0$ at
momentum transfers of $q \gtrsim 2$ fm$^{-1}$ and are found necessary to
achieve a close agreement with the experiment \cite{WiringaS98}.
%
\begin{figure*}[th]
\includegraphics[width=1\textwidth]{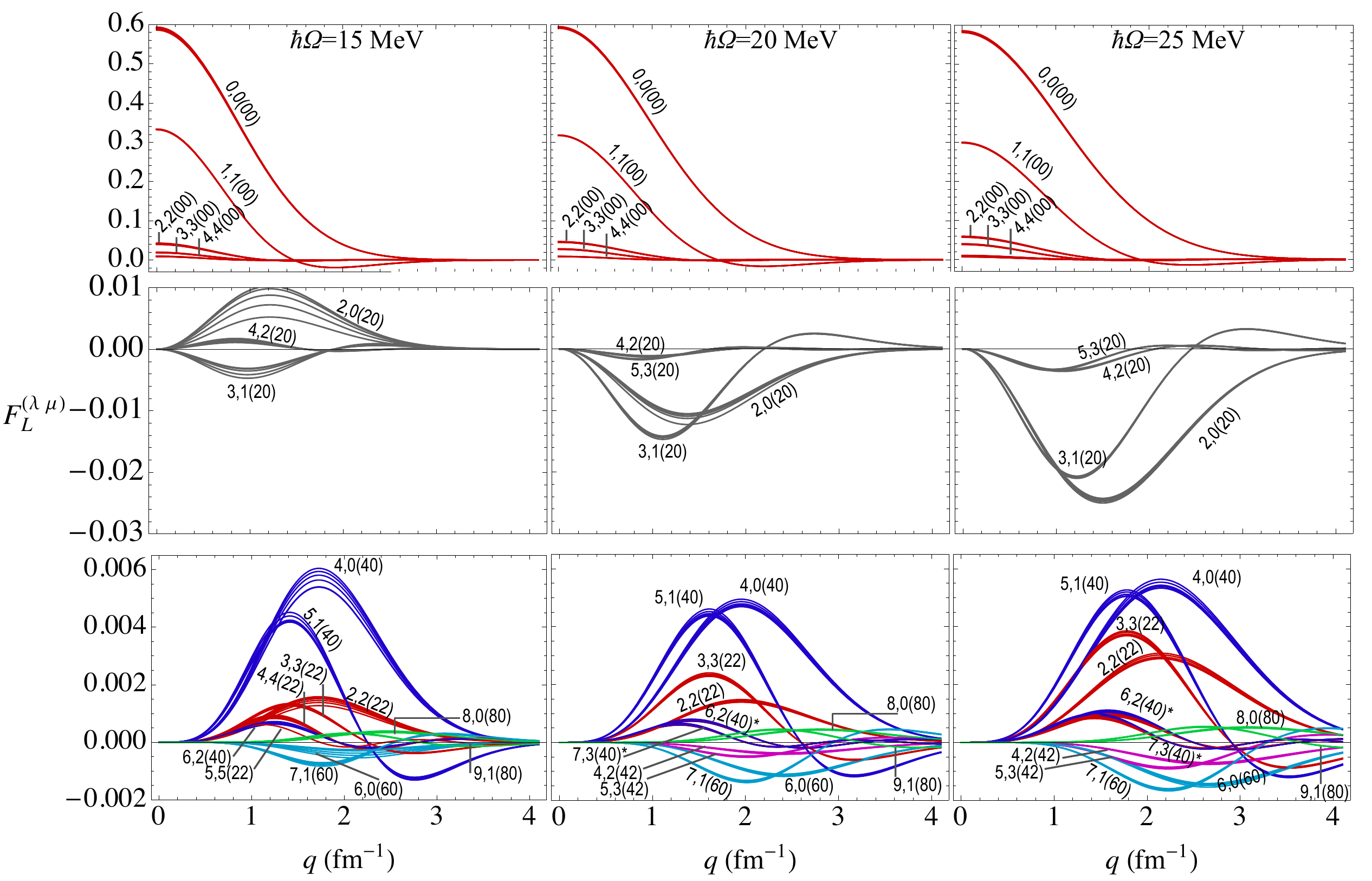} 
\caption
{
(Color online) Most dominant \SU{3} contributions  to the translationally invariant $F_L$, denoted as $F_L^{(\lambda\,\mu)}$ , for the longitudinal  $C0$ form factor, labeled by $n_1 n_2(\lambda\,\mu)$ for a particle-hole $(n_1)^1n_2^{-1}$ configuration. The $N_{\max}=12$ SA-NCSM $1^{+}$ ground  state of
	$^{6}$Li is calculated with the JISP16 bare interaction and for $\hbar\Omega=15$ MeV (left),  $20$ MeV (center), and $25$ MeV (right). Note that the vertical axis scale is reduced by an order of magnitude from the top to the bottom panels. Results are very similar to the ones obtained with the NNLO$_{\rm opt}$ bare interaction.\newline
	\footnotesize{$^*$Components $6,2(4\,0)$ and  $7,3(4\,0)$ lie almost on the top of the $4,4(2\,2)$ and  $5,5(2\,2)$ components, respectively, for all $q$.}
}
\label{fig:contributions_to_FL}
\end{figure*}
\begin{figure}[th]
\hspace{0.1in}\includegraphics[width=0.45\textwidth]{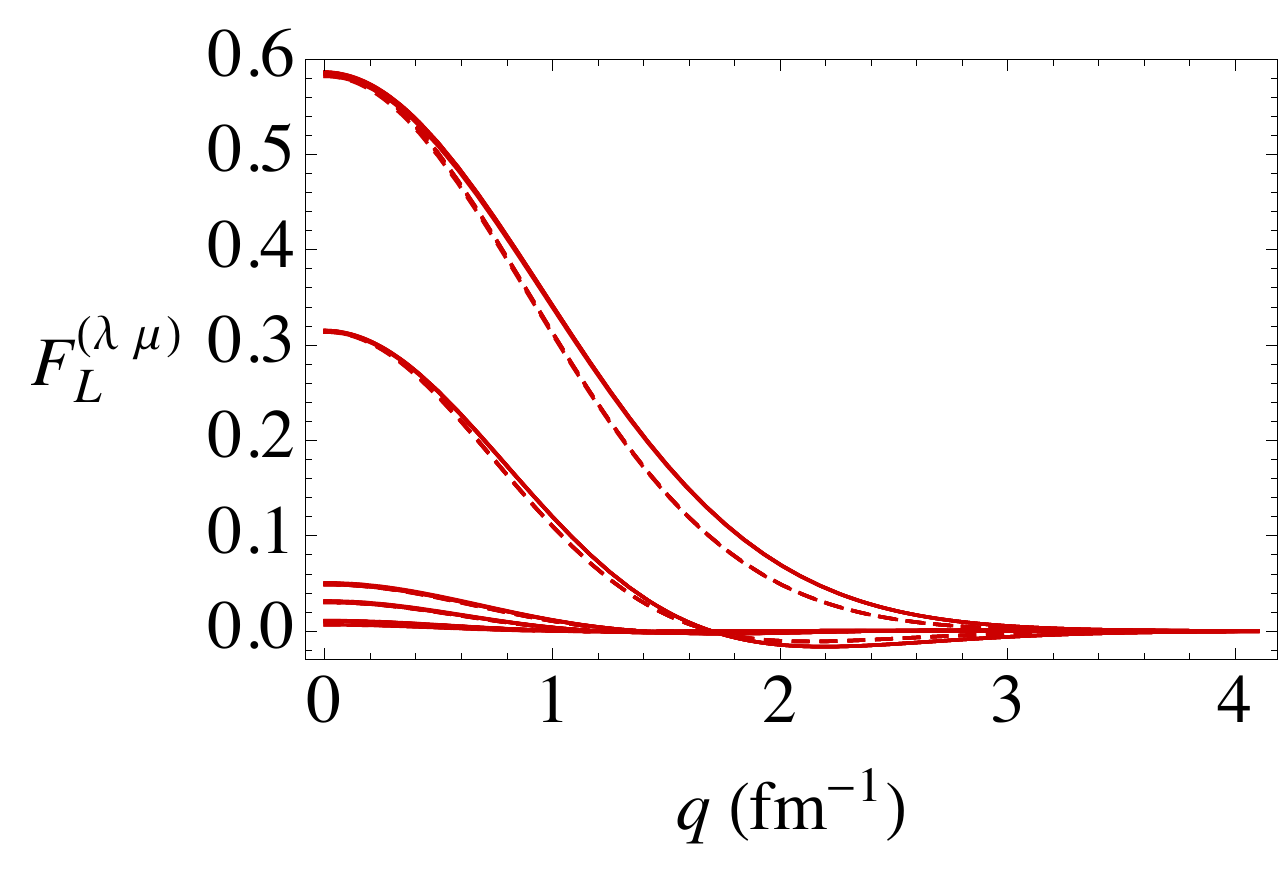} 
\vspace{0.2in}\\
\includegraphics[width=0.48\textwidth]{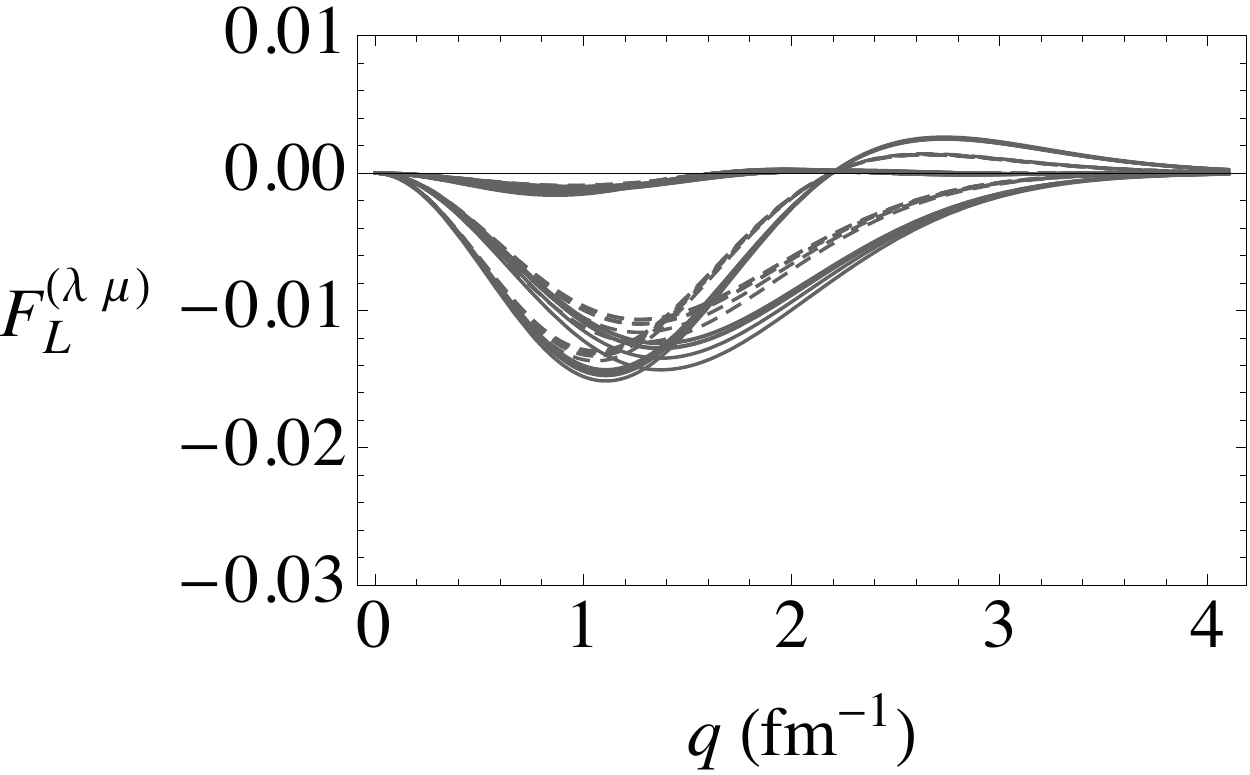} 
\vspace{0.1in}\\
\includegraphics[width=0.48\textwidth]{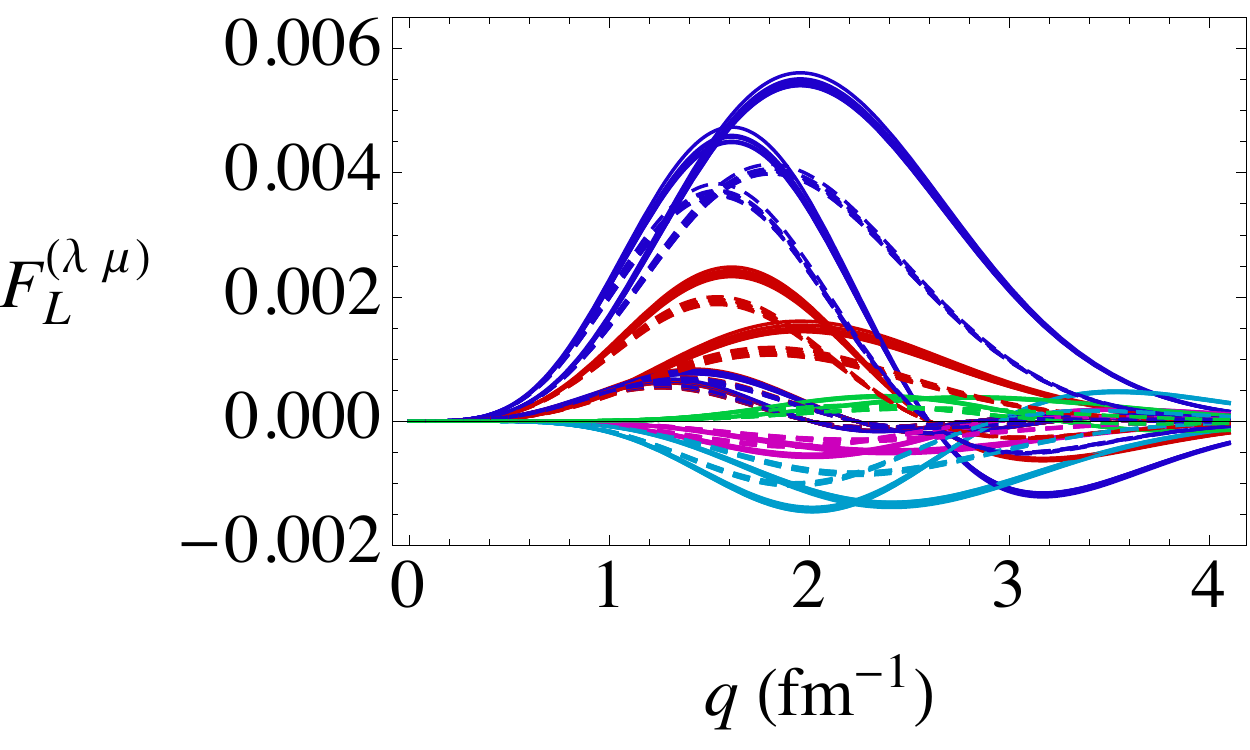} 
\caption
{
(Color online) Same as Fig. \ref{fig:contributions_to_FL}, but for  the NNLO$_{\rm opt}$ bare interaction and for $\hbar\Omega=20$ MeV, with (solid) and without (dashed) removing the CM  contribution (due to the $0s$ CM component of the wavefunctions used to calculate the OBDMEs). See Fig. \ref{fig:contributions_to_FL} for curve labeling.
}
\label{fig:contributions_to_FLcmpCM}
\end{figure}

\subsection{Important contributions to  form factors}
\subsubsection{One-body density for the ground state of $^{6}$Li}
We study the most dominant OBDMEs for the ground  state of $^{6}$Li, as they are expected to provide important contributions to the form factor. By calculating the \SU{3}-coupled OBDMEs (Fig. \ref{fig:obdmes}), the largest matrix elements are found to belong to the $n_1^1n_2^{-1}(\lambda\,\mu)= 0^10^{-1}(0\,0)$ configuration (transitions within the $s$ shell) followed by the $1^11^{-1}(\lambda\,\mu)=(0\,0)$ configuration (transitions within the $p$ shell),  or $\Delta n=|n_1-n_2|=0\ho$ transitions. Typically, all the $0\ho(0\,0)$ contributions are important together with 2\ho~$(\lambda\,\mu)=(2\,0)/(0\,2)$ and 4\ho~$(\lambda\,\mu)=(4\,0)/(0\,4)$, while there are smaller but nonnegligible components for 6\ho~$(\lambda\,\mu)=(6\,0)/(0\,6)$ and 0\ho(2\,2) (Fig. \ref{fig:obdmes}), followed by 8\ho~$(\lambda\,\mu)=(8\,0)/(0\,8)$ and 2\ho(4\,2) (not shown in the figure). 

The dominance of $k\ho(k\,0)/(0\,k)$, $k=2,4,\dots$, in $N_{\rm max}=12$ complete-space OBDMEs (as shown in Fig.  \ref{fig:obdmes}) can be recognized as another signature of the \SpR{3} symmetry, as 2\ho(2\,0) single-particle
excitations [and the conjugate (0\,2)] are described by generators of \SpR{3}, while  a stretched  coupling of such excitations yields the multiples thereof [4\ho(4\,0), 6\ho(6\,0), etc.]. These, coupled to symplectic transitions of a particle two shells down, $2\ho(0\,2)$, can also yield 0\ho(2\,2), the result of $(2\,0)\times(0\,2)$, and 2\ho(4\,2), the result of  $(4\,0)\times(0\,2)$. 

In addition, we examine the dependence of the calculated OBDMEs  on  the symmetry-based space selection (error bars in Fig. \ref{fig:obdmes}). Specifically, the deviations are defined by the range from the lowest value to the largest value of each OBDME over the set of \SU{3}-selected spaces, and are found to be reasonably small. Clearly, there is a very slight dependence on the model-space selection and on the value of \ho, with the exception of 2\ho(2\,0)/(0\,2), which is about an order of magnitude smaller than the main (0\,0) component.  However, the comparatively larger uncertainties in 2\ho(2\,0)/(0\,2) are the reason, as also shown below, for the wider spread observed in Fig. \ref{FLC0_Li6} of the selected-space $F^2_L$  for momenta above $q \gtrsim 2$ fm$^{-1}$. 
Moreover, OBDME amplitudes for $2^10^{-1}(2\,0)$ and $4^12^{-1}(2\,0)$ (and conjugates) are found to decrease for smaller \ho~values, eventually changing their sign. The observed opposite sign for small \ho~ has been suggested in  Ref. \cite{HayesK13}  based on $F_L^2$ and charge densities of $^{6}$Li and $^{12}$C for $\ho=11 - 15$ MeV. However, this effect appears to be independent of the type of the interactions employed, namely, the present study uses bare JISP16 and chiral interactions, while Lee-Suzuki effective interactions for CD-Bonn and AV$8'$ (plus a 3-body interaction) are explored in Ref. \cite{HayesK13}.
\begin{figure}[th]
\includegraphics[width=0.44\textwidth]{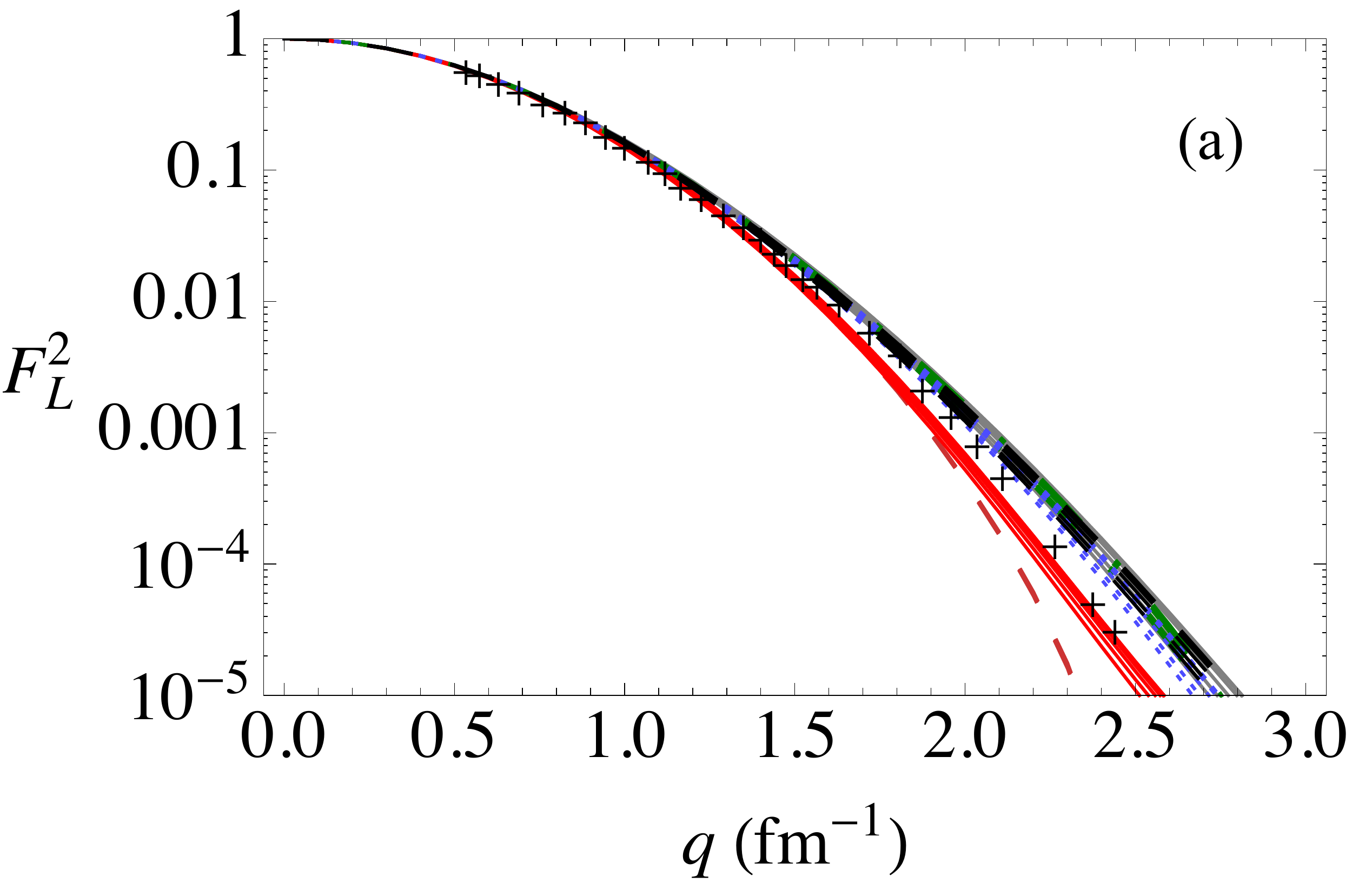} \\
\includegraphics[width=0.44\textwidth]{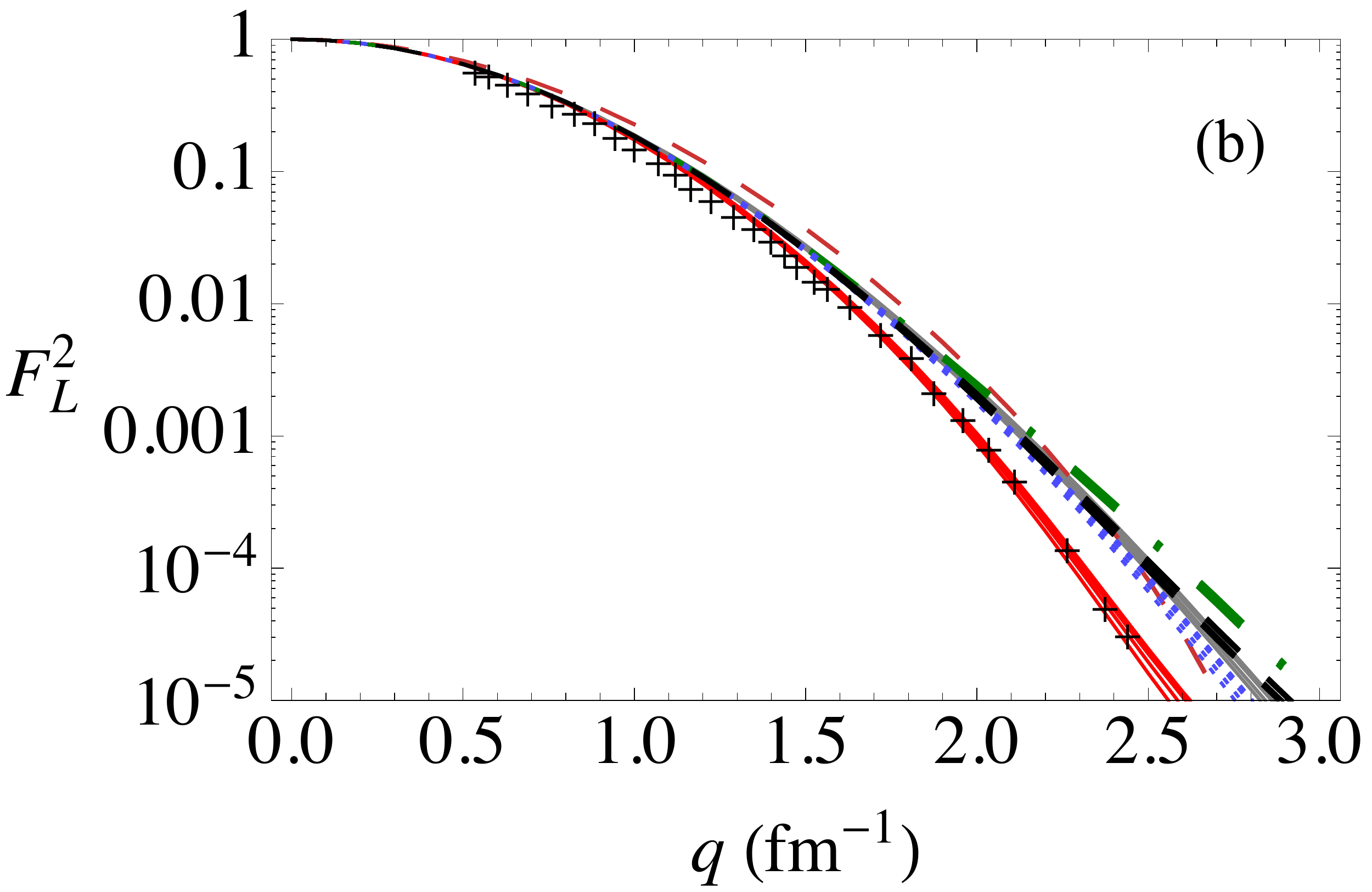} \\
\includegraphics[width=0.44\textwidth]{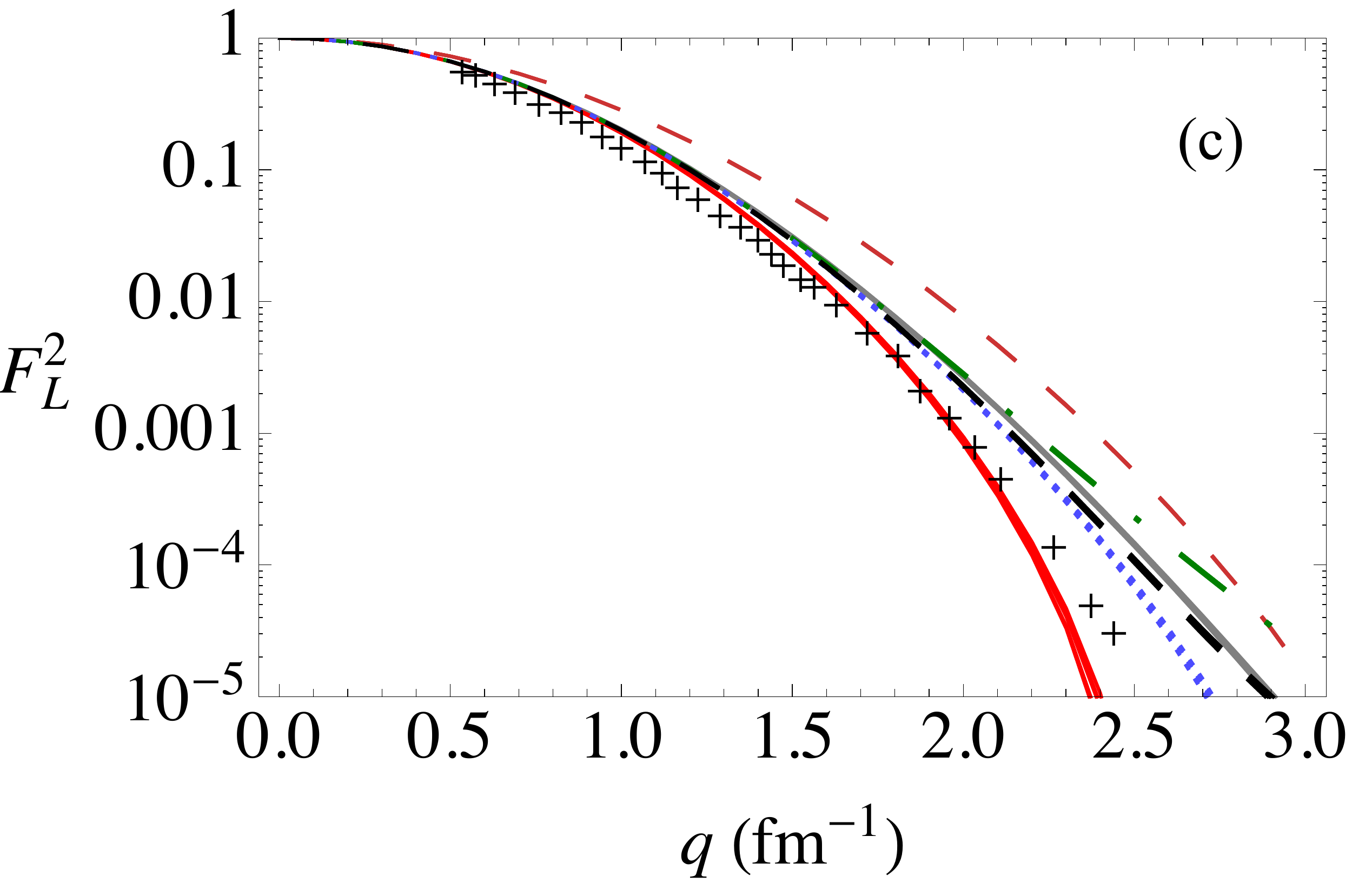} \\
\caption
{
(Color online) Longitudinal $C0$  translationally invariant form factors, constructed from a subset of selected $(\lambda\,\mu)$ OBDMEs, for the SA-NCSM $1^{+}$ $gs$ of $^{6}$Li calculated  with the bare NNLO$_{\rm opt}$ interaction (similarly for JISP16) for (a)  $\hbar\Omega=15$ MeV, (b)  $\hbar\Omega=20$ MeV, and (c)  $\hbar\Omega=25$ MeV, and compared to experiment (``$+$") \cite{LiSWY71}. The $(\lambda\,\mu)$ components [and $(\mu\, \lambda)$] included are: all (grey solid), $(0\,0)$ (red dashed), $(0\,0)+(2\,0)$ (red solid), $(0\,0)+(2\,0)+(4\,0)$ 
(green dot-dashed), $(0\,0)+(2\,0)+(4\,0)+(6\,0)$ (blue dotted), and $(0\,0)+(2\,0)+(4\,0)+(6\,0)+(8\,0)$ (black long-dashed). Deviations due to the \SU{3}-based space selection are indicated by the curve thickness. 
}
\label{fig:selectedFL}
\end{figure}

\subsubsection{Momentum dependence of the form-factor \SU{3} components}
Following Ref. \cite{RochfordD92,EscherD99}, where form factors are calculated in terms of \SU{3}-coupled OBDMEs, we can study the \SU{3} content of the corresponding electromagnetic operators and the contribution of each $(\lambda\,\mu)$ term, $F_L^{(\lambda\,\mu)}$,  to the  longitudinal  form factor of the ground  state of $^{6}$Li as a function of the momentum transfer (Fig. \ref{fig:contributions_to_FL}).  The CM-free total  longitudinal $C0$ form factor, $F^2_L$, is given by the squared sum of all such terms, $F^2_L(q)=|\sum_{n_1n_2 \lambda \mu} F_{L;n_1n_2}^{( \lambda\,\mu)}(q)|^2$.

The results show that the largest contribution for all $q$ values comes from the $(\lambda\,\mu)=(0\,0)$ (transitions within the $s$, $p$, $sd$, and $pf$ shells), spreading to larger momenta for higher \ho~(Fig. \ref{fig:contributions_to_FL}, top panels). As in the case of the OBDMEs discussed above, in addition to the strong $0\ho(0\,0)$ contribution,  the next important contribution comes from the $2\ho(2\,0)$ component ($2^10^{-1}$ and $3^11^{-1}$), which peaks around $1$-$1.5$ fm$^{-1}$ (see Fig. \ref{fig:contributions_to_FL}, middle panels, shown for a vertical axis scale an order of magnitude smaller than the one in the top panels).

For intermediate-momentum transfers, for $q\gtrsim 2$ fm$^{-1}$, $F_L$ is predominantly influenced by  $0\ho(0\,0)$, $2\ho(2\,0)$ ($2^10^{-1}$ and $3^11^{-1}$), $4\ho(4\,0)$ ($4^10^{-1}$ and $5^11^{-1}$), followed by $0\ho(2\,2)$ ($2^12^{-1}$ and $3^13^{-1}$) (Fig. \ref{fig:contributions_to_FL}, bottom panels). Compared to these contributions, the $6\ho(6\,0)$, $2\ho(4\,2)$, and $8\ho(8\,0)$ components have a peak smaller in magnitude but located at slightly higher momenta, $2.5$-$3$ fm$^{-1}$, and become comparable in their contribution around  $q\sim 3$ fm$^{-1}$.

Furthermore, the changes associated with the \SU{3}-based selection of the model space only appear to be significant for $2^10^{-1}(2\,0)$, and then only for low \ho~ (Fig. \ref{fig:contributions_to_FL}, widely spread curves). None of the other $(\lambda\,\mu)$ contributions to the $F_L$ are altered significantly by the \SU{3}-based space reduction.
In addition, for \ho=15 MeV, a slight  dependence on the space selection is observed for $3^11^{-1}(2\,0)$ as well as (but less importantly) for $4^10^{-1}(4\,0)$, $5^11^{-1}(4\,0)$,   $2^12^{-1}(2\,2)$, and  $6^10^{-1}(6\,0)$, up to $q\lesssim 2$ fm$^{-1}$. However, for all \ho, the deviation observed for $(4\,0)$ is  at least an order of magnitude smaller than that for $(2\,0)$. 

The effect of the CM $0s$ component of the SA-NCSM wavefunctions on the form factor is illustrated in Fig. \ref{fig:contributions_to_FLcmpCM}. 
As expected, the CM component suppresses the form factor due to its smearing of the translationally invariant charge density distribution.  This effect has been demonstrated, for example, in $^{6}$He \cite{Navratil04}  and $^{7}$Li  \cite{CockrellVM12}.


Finally, we consider form factors that are constructed of only several $(\lambda\,\mu)$ contributions [together with their conjugates $(\mu\, \lambda)$]: starting with a form factor constructed of the (0\,0) component only, and then consecutively adding the (2\,0), (4\,0), up to (8\,0) components (Fig. \ref{fig:selectedFL}). Clearly, the (0\,0) component makes up the predominant part of the form factor. It is interesting to note that, in this case, there is no dependence on the space selection for any \ho~ and for all $q$ values. Also, for all \ho, the addition of the (2\,0) component is found sufficient to reproduce the low-momentum regime of $F^2_L$. Except for low \ho, the (0\,0)+(2\,0) form factor decreases at intermediate $q$ values, while the consecutive addition of the (4\,0),  (6\,0), and (8\,0) components result first in an increase and then in a decrease of the intermediate-momentum $F^2_L$ (Fig. \ref{fig:selectedFL}, green dot-dashed, blue dotted, and black long-dashed curves, respectively). Those components are found to contribute the most to the  form factor. In addition, for $q \gtrsim 2.5$  fm$^{-1}$, including (2\,2) to the $F^2_L$ constructed of (0\,0),(2\,0), and (4\,0), results in a slight increase of $F^2_L$; similarly, including (4\,2) to the set of (0\,0),(2\,0), (4\,0), (2\,2), and (6\,0) results in a slight decrease of $F^2_L$. These (2\,2)  and (4\,2)  components slightly change the total $F^2_L$ and are found to be of a secondary importance. In short, for reasonable \ho~values ($>15$ MeV), the $(2\,0)$ component leads to a decrease in the intermediate-$q$ part of $F^2_L$, bringing its value closer to the experimental data, while the $(4\,0)$,  $(2\,2)$, and $(8\,0)$ are found to be foremost responsible to increase $F^2_L$  at intermediate momenta. 

We note that for smaller \ho~ values ($\lesssim 15$ MeV), 
results (Fig. \ref{fig:selectedFL}a) are in agreement with the findings of Ref. \cite{HayesK13}. Namely, the  comparatively large $(2\,0)$/$(0\,2)$ OBDME amplitudes in the wave functions, as discussed above, are found with the opposite sign to that needed to decrease $F_L^2$ and to reproduce the shape of the ($e$, $e'$) form factors together with charge radii (the relation between the two observables can be seen from the low-$q$ expansion, $F_L(q^2) \approx 1-\langle r_{\rm charge}^2 \rangle \frac{q^2}{6}+\dots$).
This has been clearly demonstrated in Fig. \ref{fig:contributions_to_FL} [the  $(2\,0)$ panel], where the case of $\ho=15$ MeV reveals  a comparative large and positive (2\,0)/(0\,2) contribution for $q>1$ fm$^{-1}$. The different behavior observed for low $\ho$ is consistent with NCSM results for the $^6$Li ground-state rms point-proton radius studied as a function of \ho~ and $N_{\max}$ using the bare JISP16 $NN$ interaction \cite{CockrellVM12}. This study has revealed that for $\ho \lesssim 15$ MeV,  the radius exhibits a larger dependence  on \ho, while a steady increase with $N_{\max}$  (implying a decrease for $F_L^2$) is observed only for $\ho >15$ MeV.
The importance of the  $(2\,0)$/$(0\,2)$ OBDME amplitudes, their $\hbar\Omega$-dependence and sign (known as `the sign problem' \cite{HayesK13}, not to be confused with the term, e.g., used in Monte Carlo approaches), merits additional investigation including their roles in other states and other nuclei \cite{LauneyHD15}.

\section{Conclusions}
Longitudinal electron scattering form factors for the ground  state of $^{6}$Li were studied in the framework of the SA-NCSM for the bare JISP16 and NNLO$_{\rm opt}$  $NN$ interactions  for a range of  $\hbar\Omega=15, 20,$ and $25$ MeV and for several \SU{3}-selected spaces, \Nmax{2}{12}, \Nmax{4}{12}, \Nmax{6}{12}, \Nmax{8}{12}, \Nmax{10}{12}, together with  the complete $N_{\max}=12$ space.  An important result is that in all cases, $\Nmax{6}{12}$ selected-space results are found to be almost identical to the $N_{\rm max}=12$ complete-space counterparts  for any momenta, shown here up to momentum transfer $q \sim 4$ fm$^{-1}$, 
while being reasonably close to experiment. This remains valid for various \ho~ values, as well as when different bare interactions are employed. Deviations in the form factor as a result of the \SU{3}-based selection of model spaces are found to decrease  for higher  \ho. This effect is more prominent for momenta $q > 2$ fm$^{-1}$. However, for high enough \ho~ values, results are almost independent from the model-space selection and, for $\ho=25$ MeV,  the $\Nmax{2}{12}$ form factor already reproduces the $N_{\rm max}=12$ complete-space result. 

The outcome shows that the largest contribution  comes from the $(\lambda\,\mu)=(0\,0)$ OBDMEs and, for all $q$ values, from the associated $(0\,0)$ contribution to the $F_L$, which makes the diagonal one-body density (within the $s$, $p$, $sd$, and $pf$ shells) most important. In addition, the $F_L$  for higher momenta, $q > 1$ fm$^{-1}$, is also influenced by   $2\ho(2\,0)$ ($2^10^{-1}$ and $3^11^{-1}$), $4\ho(4\,0)$ ($4^10^{-1}$ and $5^11^{-1}$), followed by $0\ho(2\,2)$ ($2^12^{-1}$ and $3^13^{-1}$) and also $6\ho(6\,0)$ ($6^10^{-1}$ and $7^11^{-1}$). There is a very slight dependence on the model-space selection and as one varies the value of \ho, with the exception of the 2\ho(2\,0)/(0\,2) component. However, the 2\ho(2\,0)/(0\,2)  OBDMEs are about an order of magnitude smaller than the those for the main (0\,0) component. In addition, for all \ho, only the (0\,0)+ (2\,0)/(0\,2) components are found sufficient to reproduce the low-momentum regime of $F^2_L$. The $(4\,0)$,  $(2\,2)$, and $(8\,0)$ components are the ones that are most responsible for larger $F^2_L$ values at intermediate momenta. The preponderance of $0\ho(0\,0)$, $2\ho(2\,0)$, \dots, and $8\ho(8\,0)$ together with  $0\ho(2\,2)$ and $2\ho(4\,2)$  (and their conjugates) in the OBDMEs as well as the associated contribution to $F_L$ can be recognized as another signature of the \SpR{3} symmetry.

In short, model-space selection  based on \SpR{3} and  \SU{3} symmetry considerations of the type we consider in the symmetry-guided concept of the SA-NCSM and that has been used to describe the low-lying structure of $^{6}$Li in Ref.  \cite{DytrychLMCDVL_PRL12}, properly treats, in addition,  the $^{6}$Li ground-state form factor for any momentum transfer (shown here up to $q\sim 4$ fm$^{-1}$). The symmetry-adapted model spaces include the important excitations to higher HO shells as seen in their significant contributions at low- and intermediate-momentum transfers.
 The outcome further confirms the utility of the SA-NCSM concept for low-lying nuclear states.
\newline

\acknowledgements
This work was supported in part by
the US NSF [OCI-0904874  and OCI-0904782], the US Department of Energy
[DE-SC0005248, DE-FG02-87ER40371, DESC0008485 (SciDAC-3/NUCLEI)], the National Energy Research Scientific Computing Center
[supported by DOE's Office of Science under Contract No. DE-AC02-05CH1123], the
Southeastern Universities Research Association, and the Czech
Science Foundation under Grant No. P202/12/2011.  This work also
benefitted from computing resources provided by Blue Waters, as well as the Louisiana Optical
Network Initiative and Louisiana State University's Center for Computation
\& Technology.  T. D., D.L., and T.O. acknowledge support from Michal Pajr and
CQK Holding. 


\end{document}